\pdfoutput=1

\documentclass[11pt]{article}

\usepackage{acl}

\usepackage{times}
\usepackage{latexsym}

\usepackage[T1]{fontenc}

\usepackage[utf8]{inputenc}

\usepackage{microtype}

\usepackage{multirow}
\usepackage{amsmath}
\usepackage{booktabs}
\usepackage{float}
\usepackage{algorithm}
\usepackage{algorithmic}
\usepackage{graphicx}
\usepackage{makecell}

\newcommand{\pg}{PGNN-EK}
\newcommand{\sast}{S-AST}

\usepackage{xcolor}

\title{A Neural Network Architecture for Program Understanding Inspired by Human Behaviors}

\author{Renyu Zhu$^{1}$\quad Lei Yuan$^{1}$\quad Xiang Li$^{1}$\thanks{\quad Corresponding Author}\quad Ming Gao$^{1}$\quad Wenyuan Cai$^{2}$ \\
        $^{1}$School of Data Science and Engineering, East China Normal University, Shanghai, China\\
        $^{2}$Shanghai Hypers Data Technology Inc., Shanghai, China\\
        \{52175100003, 51205903063\}@stu.ecnu.edu.cn \quad \{xiangli, mgao\}@dase.ecnu.edu.cn\\
        wenyuan.cai@hypers.com}

\begin{document}
\maketitle
\begin{abstract}
Program understanding is a fundamental task in program language processing.
Despite the success,
existing works fail to take human behaviors as reference in understanding programs. 
In this paper,
we consider human behaviors and propose the \pg\ model that consists of two main components.
On the one hand,
inspired by the ``divide-and-conquer'' reading behaviors of humans, 
we present a partitioning-based graph neural network model PGNN on the upgraded AST of codes.
On the other hand,
to characterize human behaviors of resorting to other resources to help code comprehension,
we transform raw codes with external knowledge and apply pre-training techniques for information extraction.
Finally,
we combine the two embeddings generated from the two components to output code embeddings.
We conduct extensive experiments to show the 
superior performance of \pg\ on the code summarization and code clone detection tasks.
In particular,
to show the generalization ability of our model,
we release a new dataset that is more challenging for code clone detection and could advance the development of the community.
Our codes and data are publicly available at \url{https://github.com/RecklessRonan/PGNN-EK}.
\end{abstract}

\section{Introduction}
The past decades have witnessed
the prosperity of programming platforms, such as \emph{Github} and \emph{Stack Overflow}.
These platforms generate massive open-source code\footnote{We interchangeably use code and program in this paper.} data that is named as ``Big Code'' in~\cite{DBLP:journals/csur/AllamanisBDS18}.
To automate the software development and maintenance, 
based on the ``Software Naturalness'' hypothesis~\cite{DBLP:journals/cacm/HindleBGS16},
natural language processing (NLP) techniques have been applied in program understanding. 
After that,
a series of downstream programming language processing (PLP) tasks can be performed, 
including code summarization~\cite{DBLP:conf/icse/ZhangW00020, DBLP:conf/acl/AhmadCRC20, DBLP:conf/iclr/LiuCXS021} and code clone detection~\cite{DBLP:conf/icse/ZhangWZ0WL19, DBLP:conf/wcre/WangLM0J20}.

Existing works for understanding programs mainly utilize 
three types of
information:
\emph{code context}, 
\emph{code structure} and \emph{external knowledge}.
Specifically,
code context refers to 
the token sequence in the code.
For code structure, 
each code can be parsed into various types of intermediate representations, 
such as AST (Abstract Syntax Tree), CFG (Control Flow Graph) and PDG (Program Dependence Graph).
These representations capture the structural information of codes.
Further,
there also exists external knowledge associated with codes, 
such as 
API documentation 
and other exemplary codes.
Despite the success,
all these models ignore considering human behaviors in reading programs.
Recently,~\cite{DBLP:journals/cacm/BengioLH21} suggest the potential futures of deep learning by comparing current AI methods with human learning abilities. 
This further prompts us to revisit program understanding:~\emph{
Can we develop a model that understands programs like humans?}




In the domain of programming education, how people understand codes is a topic that has been studied.
For example,
based on knowledge base including syntactical knowledge (e.g., programming basics) and semantic knowledge (e.g., API documentation),
\cite{DBLP:conf/iticse/SchulteCTBP10} offer a \emph{bottom-up} reading technique, 
which assumes that people begin with individual code lines and chunks,
and then combine them into higher-level abstractions.
Further,
\cite{DBLP:conf/iticse/ParkKCLF16} state that 
when people read codes,
reasoning about the hierarchical relationship of blocks, statements, expressions and variables is necessary. 
Based on these studies,
we conclude three key points for human understanding codes.
First,
the transition of defined variables has to be traced.
Second,
humans usually adopt a ``divide-and-conquer'' strategy,
which divides codes based on statements and then understands codes from a local-to-global view. 
Third, humans resort to external knowledge to comprehend codes, such as API documentation and code examples written by experts.



In this paper,
inspired by 
human behaviors
for code comprehension,
we propose a novel \textbf{P}artitioning-based \textbf{G}raph \textbf{N}eural \textbf{N}etwork with \textbf{E}xternal \textbf{K}nowledge (\pg). 
To capture code context and structure,
\pg\ upgrades the traditional AST 
and defines a novel subtoken-based AST called S-AST.
In S-AST,
we add edges between variables to trace the variable transitions,
edges between adjacent tree leaves from left to right
to enrich the context and structure information,
and edges between sub-nodes corresponding to subtokens tokenized from user-defined identifiers
to handle the Out of Vocabulary (OOV) problem~\cite{DBLP:conf/icse/KarampatsisBRSJ20}.
Details will be illustrated later.
After that, 
we first apply graph neural network (GNN) models on the S-AST to derive a code embedding.
To further implement the ``divide-and-conquer'' reading strategy,
we partition the S-AST into multiple subgraphs, 
which follow the sequence of statements in the original code. 
For each subgraph, 
we use GNN models to generate the subgraph embedding.
Then, these subgraph embeddings are fused to generate another code embedding.
For these two code embeddings,
since 
they are both derived from S-AST,
we further aggregate them.
On the other hand,
to characterize the dependence on external knowledge for code comprehension,
we traverse the AST of the original code to derive a sequence of tokens 
for syntactic knowledge
and then 
add the API descriptions to the end for semantic knowledge. 
We then apply CodeBERT~\cite{DBLP:conf/emnlp/FengGTDFGS0LJZ20} on the token sequence
to capture external knowledge. 
Finally, \pg\ generates the output code embedding 
by combining the embedding derived from \sast\ and the one from external knowledge.

To evaluate the model performance, we conduct experiments on the code summarization task and code clone detection task, respectively.
Before we apply \pg\ on the code clone detection benchmarks in CodeXGLUE~\cite{DBLP:journals/corr/abs-2107-07112} extracted from the 
BigCloneBench 2014 dataset~\cite{DBLP:conf/icsm/SvajlenkoIKRM14},
we notice from the leaderboard\footnote{\url{https://microsoft.github.io/CodeXGLUE/}} that the results are incredibly high, 
where the minimum F1 score is $0.949$. 
Then we dive into the characteristics of the dataset and find that the functionalities of codes in the test set have all appeared in the training set.
Therefore, the dataset is very simple.
To further test the model's generalization ability,
we construct a new dataset, where 
the test set contains codes whose functionality has never appeared in the training set.
This new dataset provides an insightful reference for further research in the community.


Our main contributions are summarized as follows:
\begin{itemize}
\item We construct a new code structure representation \sast\ that can be used to handle the OOV problem in PLP.

    \item We follow human behaviors in understanding codes and propose a novel model \pg\ that leverages code context, structure and external knowledge. Specifically,
    we put forward a novel partitioning-based graph neural network model that can effectively use code context and structure. We also present a code transformation method to utilize external knowledge in boosting comprehension.
    
    \item 
    We conduct extensive experiments on code summarization and code clone detection tasks to demonstrate the effectiveness of our model.
    In particular,
    we identify the limitation of a benchmark dataset for code clone detection and release a new dataset that is more challenging.
\end{itemize}

\begin{figure*}[t]
    \centering
    \includegraphics[width=0.95\textwidth]{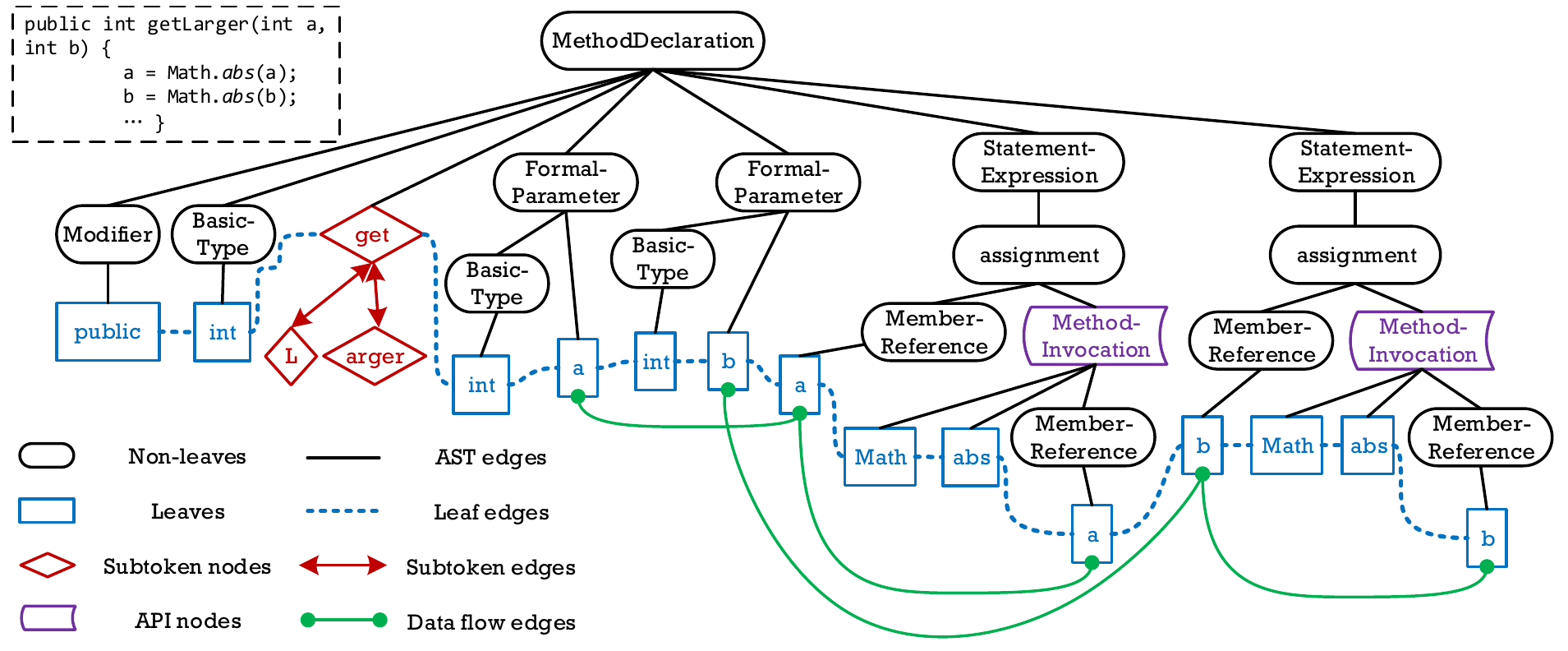}
    \caption{An example of \sast. To simplify the graph, we create a code snippet (top left), whose variables are defined with only one character, such as ``a'' and ``b''. In real tasks, the codes are longer and user-defined identifiers are more semantically complex. 
    This could add more subtoken nodes and edges.
    The figure is better viewed in color.}
    \label{fig:sub-ast}
\end{figure*}

\section{Related Work}
\subsection{Program Understanding}
Program understanding is a topic that has received wide attention.
Early works 
use either code context or structure information.
For example,
taking codes as raw texts,
some works use 
language models~\cite{DBLP:conf/pldi/RaychevVY14, DBLP:conf/sigsoft/AllamanisBBS15}, RNN-series~\cite{DBLP:journals/corr/ZarembaS14, DBLP:journals/corr/DamTP16}
and attention~\cite{DBLP:conf/acl/IyerKCZ16} to represent codes.
However,
different from natural language, 
programs are more structural,
which can be parsed into intermediate graphs, 
such as AST. 
Many works for code analysis are then proposed based on AST,
such as AST-based LSTM~\cite{DBLP:conf/ijcai/WeiL17}, AST-based CNN~\cite{DBLP:conf/iwpc/YuLCLXW19}, ASTNN~\cite{DBLP:conf/icse/ZhangWZ0WL19}, code2vec~\cite{DBLP:journals/pacmpl/AlonZLY19}, and code2seq~\cite{DBLP:conf/iclr/AlonBLY19}. 
Recently,
GNN models have also been applied in code understanding.
Since the original AST is actually a tree that is sparse,
these works~\cite{DBLP:conf/iclr/AllamanisBK18,DBLP:conf/wcre/WangLM0J20,DBLP:conf/aaai/WangL21a}
first add edges to AST to make it more connected and then apply GNN models.
Further,
there are also works~\cite{DBLP:conf/nips/YuZW0NW20, DBLP:conf/icml/CumminsFBHOL21, DBLP:conf/iclr/LiuCXS021} that utilize other intermediate graphs such as
CFG, PDG and CPG~\cite{DBLP:conf/sp/YamaguchiGAR14}.
Recently, 
approaches that use both code 
context and structure are proposed.
For example,
\citet{DBLP:conf/iclr/HellendoornSSMB20}
and~\citet{DBLP:conf/iclr/ZugnerKCLG21}
incorporate the structure information derived from AST, such as edge weights and node distances, into the context attention computation in Transformer~\cite{DBLP:conf/nips/VaswaniSPUJGKP17}. 


Despite the success,
all these methods
only consider the code context and structure information.
There are also approaches that utilize the external knowledge associated with codes.
For example,
some methods 
apply pre-training techniques in NLP to boost comprehension, such as CodeBERT~\cite{DBLP:conf/emnlp/FengGTDFGS0LJZ20}, GPT-C~\cite{DBLP:conf/sigsoft/SvyatkovskiyDFS20} and PLBART~\cite{DBLP:conf/naacl/AhmadCRC21}.
There are also works 
that incorporate code characteristics into pre-training models, such as GraphCodeBERT~\cite{DBLP:conf/icml/PengZLKHL21}, 
OSCAR~\cite{DBLP:conf/icml/PengZLKHL21} 
and InferCode~\cite{DBLP:conf/icse/BuiYJ21}. 
Further,
API is another external source for program understanding,
which has been introduced in many works~\cite{DBLP:conf/ijcai/HuLXLLJ18, DBLP:conf/acl/XuJYVN20}. 
However, all these methods ignore considering human behaviors in program understanding.



\subsection{Code Summarization and Code Clone Detection}
In this paper, we focus on two program understanding downstream tasks: code summarization and code clone detection.
For code summarization, 
some works~\cite{DBLP:conf/acl/IyerKCZ16, DBLP:conf/acl/AhmadCRC20} use code context only,
some methods~\cite{DBLP:conf/icse/LeClairJM19, DBLP:conf/iclr/AlonBLY19} use code structure only, while there are also models~\cite{DBLP:conf/iclr/HellendoornSSMB20, DBLP:conf/iclr/ZugnerKCLG21} that use both information.
Further,
\citet{DBLP:conf/iclr/LiuCXS021} introduce 
external knowledge for performance improvement.
For code clone detection, 
existing works mainly employ code structure~\cite{DBLP:conf/ijcai/WeiL17, DBLP:conf/icse/ZhangWZ0WL19, DBLP:conf/wcre/WangLM0J20} and pre-training models~\cite{DBLP:conf/emnlp/FengGTDFGS0LJZ20, DBLP:conf/naacl/AhmadCRC21}. 


\section{\sast\ Construction}

In this section, we construct \sast. 
The original AST has two main limitations:
\begin{itemize}
    \item \textbf{Low connectivity}. 
    The original AST is actually tree-structured,
    where every two nodes are minimally connected with only one path.
    This could lead to a long distance between leaf nodes. As pointed out in~\cite{DBLP:conf/iclr/0002Y21},
    directly applying GNN models in tree-shaped graphs could cause the long-range problem.
    \item \textbf{OOV problem}. 
    User-defined identifiers in codes can be arbitrarily complex and most of them are compound words,
    which could induce a large vocabulary size.
    For example, 
    the training set size in the benchmark dataset CodeXGLUE~\cite{DBLP:journals/corr/abs-2102-04664} for code summarization is $164,814$, while the vocabulary size for AST nodes is $620,256$. 
    After we split the nodes by camel case and underscores~\cite{DBLP:conf/icml/CvitkovicSA19}, 
    the vocabulary size is still as high as $201,286$. A very large vocabulary could cause the OOV problem~\cite{DBLP:conf/acl/JeanCMB15} and thus adversely affect the model performance.
\end{itemize}

To improve the connectivity of the AST,
there exist some works~\cite{DBLP:conf/iclr/AllamanisBK18, DBLP:conf/wcre/WangLM0J20, DBLP:conf/aaai/WangL21a} that 
add edges to the AST.
However,
these methods cannot address the OOV problem. 
Therefore,
we propose a new code intermediate graph \sast,
as shown in Figure~\ref{fig:sub-ast}.
Similar as in~\cite{DBLP:conf/iclr/AllamanisBK18, DBLP:conf/wcre/WangLM0J20},
we add data flow edges to trace variable transitions
and connect adjacent leaf nodes to encourage learning from contexts.
To solve the OOV problem,
we further reduce the vocabulary size by
using the tokenizer of RoBERTa~\cite{DBLP:journals/corr/abs-1907-11692} 
to tokenize every leaf node in the AST. 
When a leaf node can be tokenized into multiple subtokens, 
we keep the first subtoken as the parent node and take other subtokens as its children. 
For example,
the token ``getLarger'' is divided into the parent node ``get'' and the children nodes ``L'' and ``arger''. 
These new parent-children connections are defined as subtoken edges. 
With these three types of edges added, 
we increase the number of edges in the AST and improve the graph connectivity. 
Further,
the vocabulary size could be significantly reduced.
In our experiments,
we use javalang\footnote{\url{https://github.com/c2nes/javalang}} to generate Java AST and reduce the vocabulary size to $50,336$, 
where $50,265$ is the size of original RoBERTa vocabulary and $71$ is the number of keywords in non-leaf nodes defined by javalang.


\section{Algorithm}

\begin{figure}[H]
    \centering
    \includegraphics[width=0.99\columnwidth]{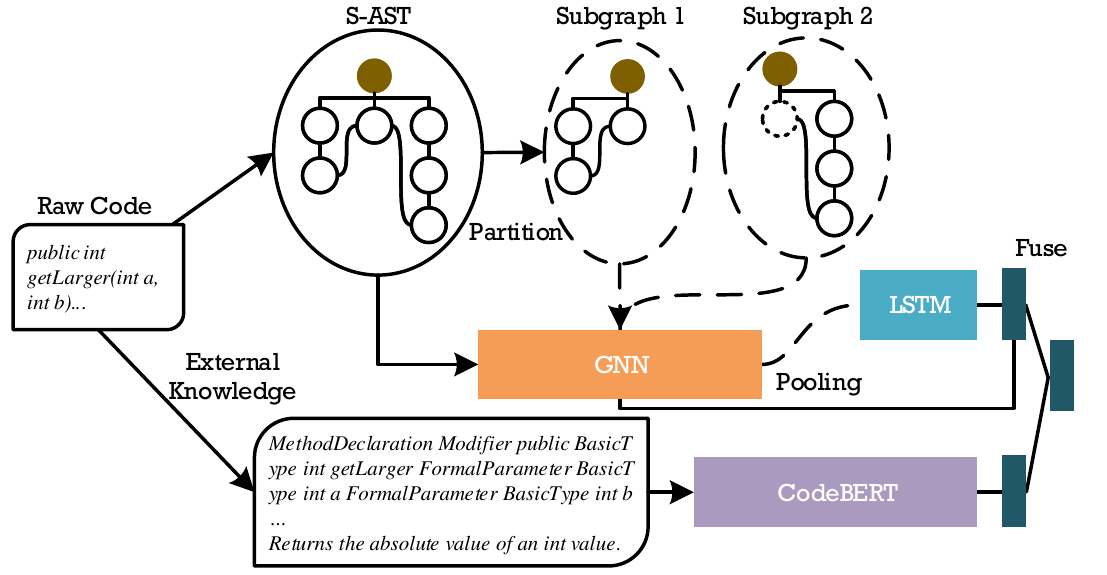}
    \caption{The overall architecture of PGNN-EK}
    \label{fig:pgnn-ek}
\end{figure}

In this section, we introduce the \pg\ model,
which
is composed of two main components. 
On the one hand,
the
partitioning-based graph neural network model (PGNN) is proposed to follow the ``divide-and-conquer'' behaviours of humans to understand programs. 
On the other hand, 
\pg\ leverages external knowledge to enhance the model's capability.
The overall architecture of \pg\ is summarized in Figure~\ref{fig:pgnn-ek}.


\subsection{Partitioning-based Graph Neural Networks}
As illustrated in~\cite{DBLP:conf/iticse/SchulteCTBP10} and~\cite{DBLP:conf/iticse/ParkKCLF16},
the bottom-up
reasoning on the hierarchical relationship of statements plays an essential role in human understanding.
Therefore,
we propose a statement-based partitioning algorithm to divide \sast\ into multiple subgraphs.
Since \sast\ is no longer a tree,
for convenience,
we first keep subtokens and their edges in-between in \sast, 
and remove edges linking variables and those connecting adjacent leaf nodes, to derive a tree structure. 
After that,
we calculate the number of nodes in each subtree of the root node
and each subtree corresponds to a statement of the raw code.
Then, 
we accumulate the number of nodes in subtrees from left to right. When the sum exceeds the pre-defined threshold $\lambda$, 
we group these subtrees into one subgraph and reset the sum to zero.
If the current subgraph is not the first one,
for each variable node in it,
we also add to the subgraph the closest node indicating the same variable in previous subgraphs to trace the variable transition.
After the subgraph is derived,
we 
add edges between 
nodes that represent the same variable
and also connect
adjacent leaf nodes as in the original S-AST. 
We repeat this process until all subtrees are visited. 
Note that if the node number of the last subgraph is smaller than $\lambda/2$,
we merge the last subgraph into the penultimate subgraph. 
Finally,
we summarize the pseudocodes of the partitioning algorithm in Alg.~\ref{alg:algorithm}. 

After subgraphs are derived,
as in~\cite{DBLP:conf/iclr/HellendoornSSMB20},
we adopt GGNN~\cite{DBLP:journals/corr/LiTBZ15} as the graph embedding model,
which uses a multi-layer perceptron (MLP) and a gated recurrent unit (GRU) to perform message passing and embedding updating.
Specifically,
at the $(l+1)$-th layer, 
to update the embedding $\mathbf{h}_{i}^{l+1}$ of node $x_i$, 
we have:
\begin{align*}
    \mathbf{m}_i^{l+1} = & \sum_{j\in \mathcal{N}_i} \text{MLP}(\mathbf{h}_{j}^{l}, \mathbf{e}_{ij}), \\
    \mathbf{h}_{i}^{l+1} = & \text{GRU}(\mathbf{m}_i^{l+1}, \mathbf{h}_{i}^{l}),
\end{align*}
where 
$\mathcal{N}_i$ is the neighbor set of $x_i$
and $\mathbf{e}_{ij}$ is the feature vector of the edge between $x_i$ and $x_j$.
After node embeddings are generated, 
we use a READOUT function to obtain the graph embedding $\mathbf{G}$:
$$
\mathbf{G} = \text{READOUT}(\{\mathbf{h}_i\}).
$$
We repeat the above process on each subgraph 
to derive 
a list of subgraph embeddings $\mathbf{L} = [\mathbf{G}_1, \mathbf{G}_2, \cdots, \mathbf{G}_n]$, 
where $n$ is the number of subgraphs. 
Next, we keep the order of the subgraph list and feed $\mathbf{L}$ into an unidirectional LSTM: 
$$
\mathbf{O} = \text{LSTM}(\mathbf{L}).
$$
Inspired by the skip connection~\cite{DBLP:conf/cvpr/HeZRS16},
we also perform GGNN on the whole S-AST graph to derive a code embedding $\mathbf{C}$.
Finally, 
we concatenate $\mathbf{C}$ and the last output $\mathbf{O}[-1]$ of LSTM.
We further feed the result into a fully connected layer to get the output code embedding $\mathbf{E}_p$:
$$
\mathbf{E}_p = \text{FC}(\text{Concat}(\mathbf{C}, \mathbf{O}[-1])).
$$

\subsection{External Knowledge}

To help understand programs,
people often resort to external knowledge.
For example,
humans usually learn from massive exemplary codes written by experts for better syntactic comprehension,
which are in the format of programming language.
Further,
API documentation is written in natural language and provides semantic details on functions.
Therefore,
a research question arises:
\emph{how to fuse these external syntactic and semantic knowledge into our model?} 


To address the problem,
we use 
pre-training techniques in 
programming language processing (PLP),
which are 
trained on massive code corpus to learn programming basics. 
In particular, we adopt CodeBERT~\cite{DBLP:conf/emnlp/FengGTDFGS0LJZ20},
which is a bimodal pre-trained model for both programming language and natural language.

\begin{figure}[t]
    \centering
    \includegraphics[width=0.95\columnwidth]{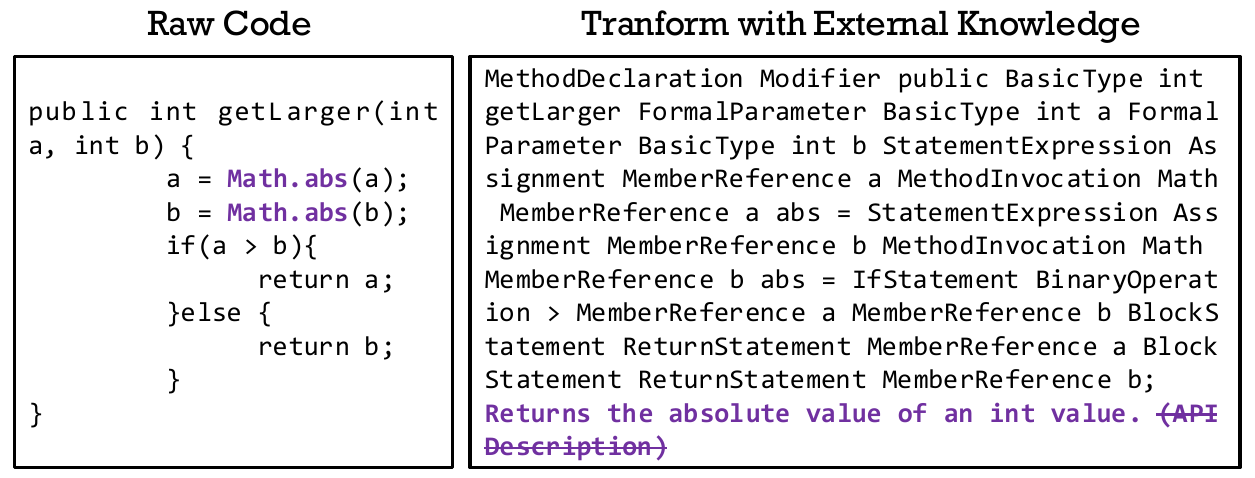}
    \caption{A toy example on code transformation with external knowledge. 
    The last sentence in the right box is the API description of \emph{Math.abs}.}
    \label{fig:transform}
\end{figure}

Before CodeBERT is applied,
we first combine the raw code and API descriptions.
To enrich the syntactic information contained in the raw code,
we perform pre-order traversal on the AST of the code to obtain a sequence of tokens and replace the raw code.
This is because 
the AST includes extra code-related information, such as statements, variables and operations. 
Then we append the corresponding API description to the end. 
 A toy example of transformation is shown in Figure~\ref{fig:transform}. 
 Finally, we feed the transformed context $\mathbf{T}$ into the pre-trained CodeBERT\footnote{\url{https://huggingface.co/microsoft/codebert-base}} and obtain the embedding $\mathbf{E}_e$:
$$
\mathbf{E}_e = \text{CodeBERT}(\mathbf{T}).
$$

Finally, 
we concatenate the output embeddings of PGNN and CodeBERT, 
and feed the result into 
a fully connected layer
to obtain the final embedding $\mathbf{E}_f$:
$$
\mathbf{E}_f = \text{FC}(\text{Concat}(\mathbf{E}_p, \mathbf{E}_e)).
$$

\section{Experiments}
In this section, we evaluate the performance of \pg. We conduct experiments on two program understanding tasks: code summarization and code clone detection. 
For each task, 
we use two benchmark datasets, 
whose statistics are listed in Table~\ref{tab:dataset_stat}. 
\begin{table*}[t]
    \centering
    \caption{The statistics of datasets}
    \resizebox{\linewidth}{!}{
    \begin{tabular}{c|c|c|c|c|c}
    \toprule
    Task & Dataset & Training & Validation & Test & Description \\
    \midrule
     \multirow{2}*{Code summarization}    & CodeSearchNet-Java (CSN) & 164,814 & 5,179 & 10,952 & Provided by CodeXGLUE\\
         & TL-CodeSum (TLC) & 69,708 & 8,714 & 8,714 & Original \\
    \midrule
      \multirow{2}*{Code clone detection} & BigCloneBench (BCB) & 901,028 & 415,416 & 415,416 & Provided by CodeXGLUE \\
     & BigCloneBench-Function (BCB-F) & 398,110 & 78,602 & 81,202 & Split by functionality \\
    \bottomrule
    \end{tabular}
    }
    \label{tab:dataset_stat}
\end{table*}

\subsection{Implementation details}
In our experiments,
we use the AdamW optimizer and linear schedule from~\cite{wolf-etal-2020-transformers}
to update model parameters. 
For fair comparison, 
we run all experiments on $2$ Tesla V$100$ with $32$G memory. 
For PGNN, we set the number of GNN layers, the number of LSTM layers, the embedding size of GNN node, and the embedding size of LSTM hidden layer to $3$, $2$, $768$ and $768$, respectively. 
We choose the mean operator as the READOUT function. 
To avoid overfitting, 
we set the dropout rate to $0.2$ in PGNN. 
We implement GNNs based on PyTorch Geometric~\cite{Fey/Lenssen/2019}. 
In the EK-enhanced component, 
we obtain $51,191$ method-description pairs 
after preprocessing the API documentation\footnote{\url{https://www.oracle.com/java/technologies/javase-jdk8-doc-downloads.html}}.
For pair examples,
see Appendix~\ref{appedix:api}.
In the code summarization task, 
we add a $6$-layer Transformer-based decoder to generate summarization as in CodeBERT. 
We set learning rate to $0.00005$, 
batch size to $16$, training steps to $50,000$, 
maximum code length to $256$ and maximum summarization length to $32$, respectively. 
In the code clone detection task, 
as suggested by ~\cite{DBLP:conf/rep4nlp/NeculoiuVR16}, we double the \pg\ to a siamese neural network 
to calculate code similarity. 
We set learning rate to $0.00005$, batch size to $4$, 
training steps to $200,000$ and maximum code length to $400$, respectively. 

\subsection{Code Summarization}
Code summarization aims at generating natural language comments for codes. We evaluate the performance of \pg\ on two benchmark datasets, 
which are TL-CodeSum (shorted as TLC)~\cite{DBLP:conf/ijcai/HuLXLLJ18} and the Java subset of CodeSearchNet (shorted as CSN)~\cite{DBLP:journals/corr/abs-1909-09436}.
For TLC, we use the original dataset. 
For CSN, 
we use the version provided by CodeXGLUE~\cite{DBLP:journals/corr/abs-2102-04664}. 
For fair comparison, 
we use the smoothed BLEU-4 score~\cite{DBLP:conf/coling/LinO04} 
as in CodeXGLUE.
The larger the score,
the better the model performance.
We compare our model with five representative baselines, including  CodeNN~\cite{DBLP:conf/acl/IyerKCZ16}, NCS~\cite{DBLP:conf/acl/AhmadCRC20}, Rencos~\cite{DBLP:conf/icse/ZhangW00020}, CodeBERT~\cite{DBLP:conf/emnlp/FengGTDFGS0LJZ20} and PLBART~\cite{DBLP:conf/naacl/AhmadCRC21}.
Due to the space limitation,
we move the details of these baselines to Appendix~\ref{appendix:baselines}.

Table~\ref{tab:res_cs} shows the code summarization results.
Note that
the results of CodeNN, NCS and Rencos are directly taken from \cite{DBLP:journals/corr/abs-2107-07112}. 
Also, the results of CodeBERT and PLBART on CSN are derived from the leaderboard of CodeXGLUE. 
For their results on TLC,
we run the codes released by the authors of the paper and set hyper-parameters according to the original paper.
From the table, 
we see that, 
due to the fusion of external knowledge,
pre-training models CodeBERT, PLBART and \pg\ outperform other models on both datasets. 
Further,
\pg\ performs the best.
The gaps between \pg\ and the runner-up model PLBART on CSN and TLC are $0.5$ and $1.05$, respectively. 
This shows the importance of considering human behaviors for code comprehension.
We also observe that 
scores on TLC are substantially larger than that on CSN.
This is because codes in the training set and the test set of TLC are considerably more similar in functionalities, which will be elaborated in the next section. 

\begin{table}[H]
    \centering
    \caption{Code summarization results. We highlight the best results in bold. * indicates that the improvements are statistically significant for $p< 0.01$ with paired t-test.}
    \begin{tabular}{c|c|c}
    \toprule
    Model &  CSN  &  TLC \\
    \midrule
    CodeNN & 8.58 & 33.03 \\
    NCS & 11.19 & 44.25 \\
    Rencos & 11.80 & 46.81\\
    \midrule
    CodeBERT & 17.65 & 48.53  \\
    PLBART & 18.45 & 50.01 \\
    \midrule
    \pg & \textbf{18.95}$^{*}$ & \textbf{51.06}$^{*}$\\
    \bottomrule
    \end{tabular}
    \label{tab:res_cs}
\end{table}

\subsection{Code Clone Detection}
\label{sec:exp_ccd}

The goal of code clone detection is to detect whether two code fragments implement the same functionality. 
Following~\cite{DBLP:conf/icse/ZhangWZ0WL19, DBLP:conf/wcre/WangLM0J20}, 
we use the BigCloneBench 2014 dataset~\cite{DBLP:conf/icsm/SvajlenkoIKRM14} and adopt the version provided by CodeXGLUE.
We short it as BCB. 

Before we apply \pg\ on BCB,
we notice from the leaderboard of CodeXGLUE that the results on BCB are incredibly high, 
where the minimum F1 score is $0.949$. 
Then we dive into the characteristics of the dataset and compare BCB with the original benchmark~\cite{DBLP:conf/icsm/SvajlenkoIKRM14}.
We find that the functionalities of codes in the test set have all appeared in the training set of BCB.
Therefore, BCB is a very simple dataset.
To test the model's generalization ability,
we construct a new dataset, named BCB-F, where 
the test set contains codes whose functionality has never appeared in the training set. 
We first extract codes from the new version benckmark~\cite{DBLP:conf/icsm/SvajlenkoR15} that has more code fragments and code functionalities. 
We next split training/validation/test set 
based on code functionalities.
Specifically, 
we construct training/validation/test set with $22/11/10$ code functionalities.
For details on the functionality splits of BCB and BCB-F,
see Appendix~\ref{appdeix:bcb-f}.
We keep the same number of positive and negative samples in all the three sets. 
The comparison between BCB and BCB-F is given in Table~\ref{tab:comp_bcb}. 
\begin{table}[H]
    \centering
    \caption{Comparisons between BCB and BCB-F}
    \resizebox{\linewidth}{!}{
    \begin{tabular}{c|c|c}
    \toprule
     &  BCB &  BCB-F \\
    \midrule
    Code fragments & 9134 & 73182 \\
    Functionalities & 10 & 43 \\
    Training/Test splitting & random sample & by functionality\\
    Ratio of positive-negative & nearly 2:1  & 1:1\\
    \bottomrule
    \end{tabular}
    }
    \label{tab:comp_bcb}
\end{table}
In addition to the pre-training models CodeBERT and PLBART,
we further compare our model with two representative methods in code clone detection, which are ASTNN~\cite{DBLP:conf/icse/ZhangWZ0WL19} and FA-AST~\cite{DBLP:conf/wcre/WangLM0J20} (For the details of these baselines,
see Appendix~\ref{appendix:baselines}).

Table~\ref{tab:res_ccd} shows the evaluation results on the two datasets. 
For BCB, we take
the results of other baseline methods from CodeXGLUE\footnote{
Specifically,
we take the results of ASTNN and FA-AST from \url{https://github.com/microsoft/CodeXGLUE/tree/main/Code-Code/Clone-detection-BigCloneBench} and that of CodeBERT and PLBART from the CodeXGLUE leaderboard. Note that PLBART only reports the F1 score on BCB.}.
For BCB-F, we run the source codes released by their authors to obtain the results.
From the table, we observe: 1) All models perform very well on BCB, indicating that the dataset is very simple. 
However, 
the best F1 score on BCB-F is only $0.724$, which shows that this dataset is very challenging.
2) The non-pre-training models ASTNN and FA-AST predict all samples to be positive and perform poorly on BCB-F, while pre-training models perform better.
This further demonstrates the importance of introducing external knowledge.
3) \pg\ achieves the best results on both datasets.
This shows that considering human behaviors in program understanding enhances the generalization ability of \pg. 

\begin{table}[H]
    \centering
    \caption{Code clone detection results w.r.t. precision (P), recall (R) and F1 measures. We highlight the best results in bold. * indicates that the improvements are statistically significant for $p< 0.01$ with paired t-test.}
    \resizebox{\linewidth}{!}{
    \begin{tabular}{c|c|c|c|c|c|c}
    \toprule
    \multirow{2}*{Model} &  \multicolumn{3}{c|}{BCB} &   \multicolumn{3}{c}{BCB-F} \\
    &  P & R &  F1 &  P & R &  F1\\
    \midrule
    ASTNN & 0.92  & 0.94 & 0.93 & 0.50 & \textbf{1.00}  & 0.67\\
    FA-AST & 0.96 & 0.94 & 0.95 & 0.50 & \textbf{1.00}  & 0.67  \\
    \midrule
    CodeBERT & 0.960 & 0.969 & 0.965 & 0.611 & 0.842 & 0.708 \\
    PLBART & - & - & 0.972 & 0.517 & 0.996  & 0.681 \\
    \midrule
    \pg\ & \textbf{0.975}$^{*}$  & \textbf{0.973}$^{*}$ & \textbf{0.974}$^{*}$  & \textbf{0.621}$^{*}$ & 0.869 & \textbf{0.724}$^{*}$\\
    \bottomrule
    \end{tabular}
    }
    \label{tab:res_ccd}
\end{table}

\subsection{Ablation Study}

\begin{table*}[t]
    \centering
    \caption{Ablation study on \pg. We highlight the best results in bold.}
    \resizebox{0.85\linewidth}{!}
    {
    \begin{tabular}{c|c|c|c|c}
    \toprule
    \multirow{2}*{Method} & CSN  & TLC  & BCB & BCB-F \\
     & (Smoothed BLEU-4) & (Smoothed BLEU-4) & (F1) & (F1) \\
    \midrule
     PGNN only  & 14.05 & 47.71 & 0.951 & 0.667\\
     EK only & 17.95 & 49.66 & 0.965 & 0.711 \\
     \pg\ with AST & 17.70 & 48.96 & 0.957 & 0.713 \\
     \pg\ without subtoken & 17.82 & 49.01 & 0.958 & 0.712 \\
     GNN-EK & 18.05 & 49.95 & 0.967 & 0.715 \\
     PGNN-CodeBERT  & 18.60 & 50.65 & 0.969 & 0.720\\
    \midrule
     \pg\ (Full Model) & \textbf{18.95} & \textbf{51.06} & \textbf{0.974}  &  \textbf{0.724} \\
    \bottomrule
    \end{tabular}
    }
    \label{tab:res_abl}
\end{table*}

We further conduct ablation study to verify the importance of its main components in \pg,
including
subtokens, the \sast\ graph, the partitioning-based GNN and the external knowledge. 
Specifically,
one variant employs only the \sast\ graph without using external knowledge. 
This helps us realize the importance of external knowledge in program understanding. 
We call this variant \textbf{PGNN only}. 
Meanwhile,
we define another variant that ignores the hierarchical relationships in code structure and uses only external knowledge.
We call this variant \textbf{EK only}. 
To further show the significance of \sast\ in code understanding, we replace \sast\ with the original AST in the variant \textbf{\pg\ with AST}.
We also implement a variant
that does not use the subtoken tokenizer to generate extra subtoken nodes and edges.
We call it \textbf{\pg\ without subtoken}.
This variant can be used to
show the importance of subtokens in addressing the OOV problem.
To show the advantage of the partitioning strategy,
we propose a variant 
\textbf{GNN-EK} that discards
the partitioning step.
Finally, we consider a variant that feeds the raw code into the pre-trained CodeBERT without transforming it with external knowledge. We call this variant \textbf{PGNN-CodeBERT}.

Table~\ref{tab:res_abl} summarizes the ablation study results.
From the table, 
we see that: 
1) \sast\ 
contains richer information than AST and can serve as an effective 
code intermediate representation in program understanding.
The introduction of subtokens nodes and edges alleviates
the OOV problem and enhances the model performance. 
2) External knowledge helps boost understanding codes.
In particular, 
code transformation with external knowledge
improves the expressiveness of the raw code.
3) The full model \pg\ outperforms other variants on all the datasets and tasks. 
This indicates the importance of every main component in \pg. It further shows that leveraging code context, code structure and external knowledge as humans is helpful for program understanding.

\subsection{The Influence of Subgraph Size}
\begin{figure}[t]
    \centering
    \includegraphics[width=0.99\columnwidth]{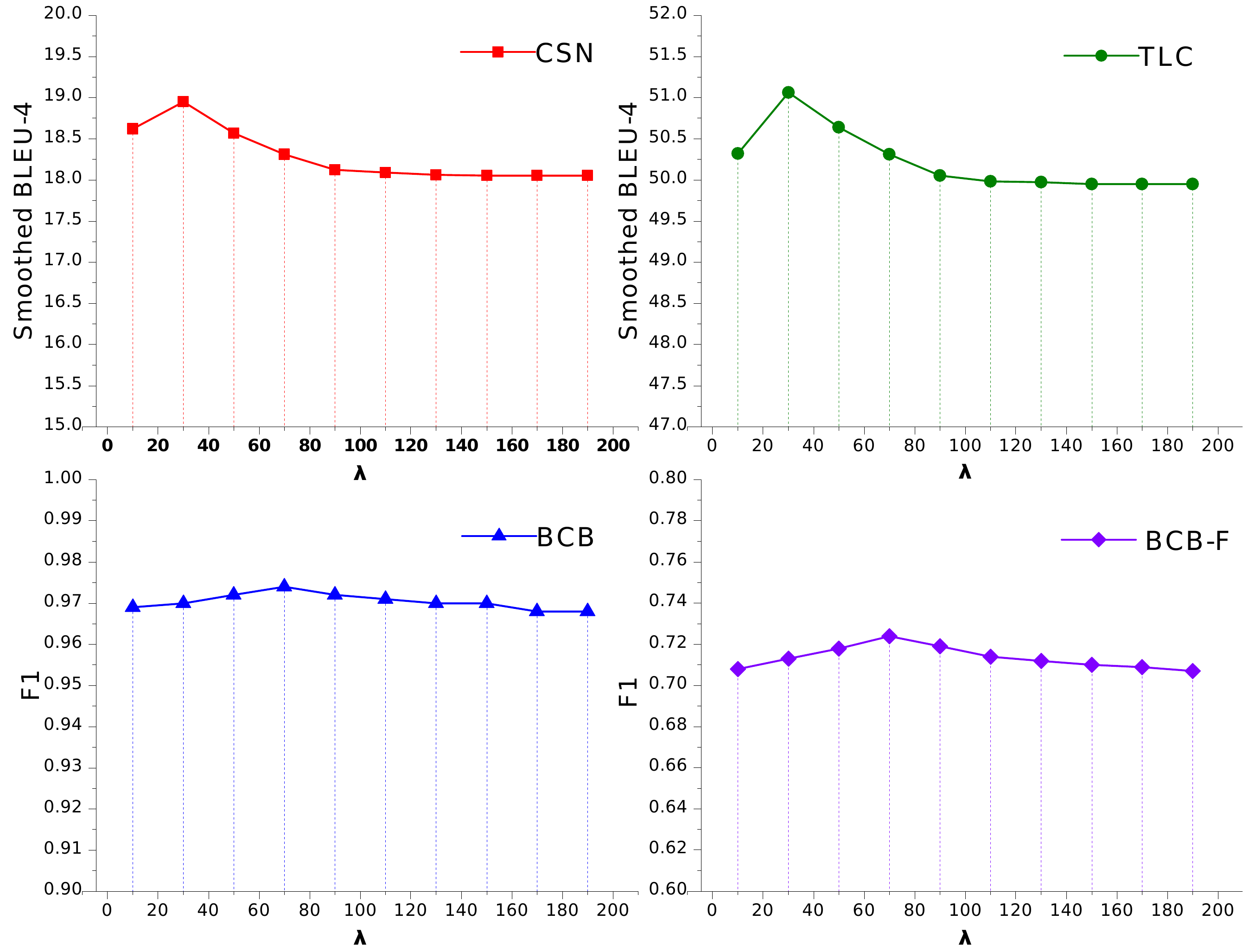}
    \caption{The influence of subgraph size on 4 datasets.}
    \label{fig:subgraph}
\end{figure}

We end this section with a hyper-parameter sensitivity analysis.
In \pg\, 
there is a key hyper-parameter $\lambda$ that is used to control the size of subgraphs. 
Here, 
we investigate the sensitivity of $\lambda$. 
We vary the value of $\lambda$ from $\{10, 30, 50, 70, 90, 110, 130, 150, 170, 190\}$, and the final prediction results of \pg\ on $4$ datasets are shown in the Figure~\ref{fig:subgraph}. 

\begin{table}[H]
    \centering
    \caption{The average number of nodes in \sast}
    \begin{tabular}{c|c|c|c|c}
    \toprule
    Datasets & CSN & TLC &  BCB &  BCB-F \\
    \midrule
    \sast\ size & 137 & 140 &  372 & 348 \\
    \bottomrule
    \end{tabular}
    \label{tab:sast_size}
\end{table}

The results indicate that 1) 
the model performance first increases and then drops, with the increase of the subgraph size.
When the subgraph size is too small, each subgraph is a code fragment that no longer represents a code statement and thus contains less information.
Further,
when the subgraph is too large,
each subgraph could be composed of statements that are of different semantic meanings,
which thus degrades the model performance.
2) 
\pg\ performs the best at $\lambda = 30$ on CSN and TLC while it achieves the best results at $\lambda = 70$ on BCB and BCB-F.
We further investigate the reason and show the average number of nodes in S-AST on the four datasets in Table~\ref{tab:sast_size}.
From the table,
BCB and BCB-F contain $\sim 2.5$ times more nodes than that in CSN and TLC.
This empirically suggests that setting $\lambda$ to be about 
$\frac{1}{5}$ to $\frac{1}{4}$ of the average node number in \sast\ could be a reasonable choice.

\section{Conclusion}
In this paper, 
we followed human understandings for programs and proposed the \pg\ model.
To enrich the code structure information and alleviate the OOV problem,
we presented the \sast\ graph based on AST,
which uses a subtoken tokenizer to generate subtoken nodes and edges between them.
Inspired by the ``divide-and-conquer'' strategy,
we proposed the partitioning-based graph neural network model on \sast\ that employs code context and structure.
To leverage the external knowledge to boost comprehension,
we transformed the raw code to fuse syntactic and semantic knowledge and utilized pre-training techniques for information extraction.
We performed extensive experiments to show the effectiveness of our model \pg\ 
on the code summarization and code clone detection tasks.
In particular,
to show the generalization ability of the model,
we released a new benchmark that is more challenging.

\section{Acknowledgments}
\label{sec:acknowledge}
This work has been supported by the National Natural Science Foundation of China under Grant No. U1911203, 
Alibaba Group through the Alibaba Innovation Research Program, 
the National Natural Science Foundation of China under Grant No. 61877018 and No.61977025,
and Shanghai Pujiang Talent Program under Grant No. 21PJ1402900.
\bibliography{acl}
\bibliographystyle{acl_natbib}

\appendix

\section{Partitioning \sast\ Algorithm}
\label{appendix:partition}
See Algorithm \ref{alg:algorithm}.
\begin{algorithm}[ht]
\caption{Partitioning \sast\ }
\label{alg:algorithm}
\textbf{Input}: A \sast\ $\mathcal{T}$ with node features $\mathcal{X}$, edge indexes $\mathcal{I}$ and edge features $\mathcal{E}$\\
\textbf{Parameter}: $\lambda$, which specifies the minimum number of nodes in the subgraph \\
\textbf{Output}: Nodes features list $\mathcal{L}_x$, edge indexes list $\mathcal{L}_i$, and edge features list $\mathcal{L}_e$ of subgraphs 

\begin{algorithmic}[1] 
\STATE Derive a tree structure $\mathcal{T}^{'}$ by removing data flow edges and adjacent leaf edges in $\mathcal{T}$;
\STATE $nodes\_sum \leftarrow 0, nodes\_set \leftarrow \{\}$;
\STATE $nf\_list, ei\_list, ef\_list, \mathcal{L}_x, \mathcal{L}_i, \mathcal{L}_e \leftarrow \{\}$;
\STATE Obtain a subtree list $\{\mathcal{S}\}$ based on subtrees of root nodes in $\mathcal{T}^{'}$ from left to right;
\FOR{$\mathcal{S}$ in $\{\mathcal{S}\}$} 
\STATE $n \leftarrow$ the number of nodes in $\mathcal{S}$;
\STATE $nodes\_sum \leftarrow nodes\_sum + n$;
\STATE Add nodes in $\mathcal{S}$ to $nodes\_set$;
\IF{$nodes\_sum \geq \lambda$ \textbf{or} $\mathcal{S}$ is the last element of $\{\mathcal{S}\}$}
\IF{$\mathcal{L}_x \neq \emptyset$}
\STATE Add closest nodes that indicate the same variables in $\mathcal{L}_x$ to $nodes\_set$ ;
\ENDIF 
\STATE Assign $nf\_list$, $ei\_list$, $ef\_list$ based on $nodes\_set$, $\mathcal{X}$, $\mathcal{I}$ and $\mathcal{E}$; 
\STATE Append $nf\_list, ei\_list, ef\_list$ to $\mathcal{L}_x, \mathcal{L}_i, \mathcal{L}_e$ respectively;
\STATE $nodes\_sum \leftarrow 0$, $nodes\_set \leftarrow \{\}$;
\ENDIF
\ENDFOR
\STATE // $A[-i]$ denotes the $i$-th element from the bottom in $A$.
\IF{size of $\mathcal{L}_x[-1] < \lambda/2$ and size of $\mathcal{L}_x > 1$}
\STATE Merge $\mathcal{L}_x[-1]$ and $\mathcal{L}_x[-2]$, $\mathcal{L}_i[-1]$ and $\mathcal{L}_i[-2]$, $\mathcal{L}_e[-1]$ and $\mathcal{L}_e[-2]$, respectively;
\ENDIF
\STATE \textbf{return}   $\mathcal{L}_x, \mathcal{L}_i, \mathcal{L}_e$
\end{algorithmic}
\end{algorithm}

\section{Examples of API-Description Pairs}
\label{appedix:api}
In the experiment. we obtain $51,191$ method description pairs after preprocessing, and Table~\ref{tab:api_example} gives some examples.

\begin{table*}[ht]
    \centering
    \caption{Examples of API-Description Pairs}
    \begin{tabular}{c|c}
    \toprule
     APIs &  Descriptions \\
    \midrule
    \emph{Math.abs} & Returns the absolute value of an int value.\\
    \emph{Arrays.hashcode} &  Returns a hash code based on the contents of the specified array.\\
    \emph{Scanner.hasNext} & Returns true if this scanner has another token in its input. \\
    \emph{Color.getRGB} & Returns the RGB value representing the color in the default sRGB ColorModel.\\
    \bottomrule
    \end{tabular}
    \label{tab:api_example}
\end{table*}

\section{Baselines Introduction}
\label{appendix:baselines}

We compare our model with five representative models in code summarization task:
\begin{itemize}
    \item CodeNN~\cite{DBLP:conf/acl/IyerKCZ16} is the first method that applies deep neural networks in code summarization. It uses a classical attention-based encoder-decoder framework from Neural Machine Translation (NMT).
    \item NCS~\cite{DBLP:conf/acl/AhmadCRC20} applies Transformer~\cite{DBLP:conf/nips/VaswaniSPUJGKP17} to model the pairwise relationship between code tokens and capture their long-term dependencies. 
    \item Rencos~\cite{DBLP:conf/icse/ZhangW00020} proposes an attention-based encoder-decoder model and enhance it with the most similar code snippets retrieved from the training set.
    \item CodeBERT~\cite{DBLP:conf/emnlp/FengGTDFGS0LJZ20} is a bimodal pre-training model for programming and natural languages based on RoBERTa~\cite{DBLP:journals/corr/abs-1907-11692}.
    \item PLBART~\cite{DBLP:conf/naacl/AhmadCRC21} is a sequence-to-sequence pre-training model based on BART~\cite{DBLP:conf/acl/LewisLGGMLSZ20}.
\end{itemize}

In addition to the pre-training models CodeBERT and PLBART, we further compare our model with two representative model in code clone detection task:
\begin{itemize}
    \item ASTNN~\cite{DBLP:conf/icse/ZhangWZ0WL19} proposes an AST-based neural network that splits AST into a sequence of statement trees and applies a bidirectional RNN model to produce source code representation.
    However, it ignores external knowledge associated with codes.
    \item FA-AST~\cite{DBLP:conf/wcre/WangLM0J20} augments original AST with explicit control and data flow edges, then introduces two different types of GNNs to detect code clones.
\end{itemize}

\section{Functionalities Splits in BCB and BCB-F}
\label{appdeix:bcb-f}
For BCB, the functionalities in Train/Val/Test set are:
\begin{itemize}
    \item \textbf{Train:} Web Download, Secure Hash(MD5), Copy a File, Decompress Zip, FTP Authenticated Login, Bubble Sort, Init. SGV with Model, SGV Selection Event Handler, Create Java Project(Eclipse), SQL Update and RollBACK.
    \item \textbf{Val:} Same to \textbf{Train}.
    \item \textbf{Test:} Same to \textbf{Train}.
\end{itemize}

For BCB-F, the functionalities in Train/Val/Test set are, where the emphasis discloses the whole $10$ functionalities that exist in BCB:
\begin{itemize}
    \item \textbf{Train:} \emph{Decompress Zip}, \emph{Copy a File}, Get Prime Factors, File Dialog, Resize Array, Get MAC Address String, Parse CSV File, \emph{Secure Hash(MD5)}, Send Email, Load Custom Font, \emph{Create Java Project(Eclipse)}, Extract Matches Using Regex, Open File in Desktop Application, Connect to Database, Load File to Byte Array, Call Method Using Reflection, Take Screenshot to File, Write PDF File, Delete Folder and Contents, Copy Directory, Binary Search, Delete Folder and Contents.
    \item \textbf{Val:} \emph{SQL Update and RollBACK}, \emph{Bubble Sort}, Execute External Process, XMPP Send Message, Zip Files, Convert Date String Format, Secure Hash, GCD, \emph{SGV Selection Event Handler}, \emph{Init. SGV with Model}, Play Sound.
    \item \textbf{Test:} Shuffle Array in Place, Create Encryption Key Files, Load Custom Font, Encrypt to File, Parse XML to DOM, CRC32 File Checksum, Transpose a Matrix, Test Palindrome, \emph{Web Download}, \emph{FTP Authenticated Login}.
\end{itemize}

\end{document}